\documentclass[journal,comsoc]{IEEEtran}

\usepackage[T1]{fontenc}

\usepackage{cite}

\ifCLASSINFOpdf
  \usepackage[pdftex]{graphicx}
\else
  \usepackage[dvips]{graphicx}
\fi
\usepackage{amsmath}
\usepackage[cmintegrals]{newtxmath}
\ifCLASSOPTIONcompsoc
  \usepackage[caption=false,labelformat=simple,font=normalsize,labelfont=sf,textfont=sf]{subfig}
\else
  \usepackage[caption=false,labelformat=simple,font=footnotesize]{subfig}
\fi

\usepackage{url}

\usepackage{cite}
\usepackage{algorithmic}
\usepackage{graphicx}
\usepackage{epstopdf}
\usepackage{textcomp}
\usepackage{xcolor}
\usepackage{diagbox}
\usepackage{multirow}

\definecolor{darkGreen}{rgb}{0.0, 0.5, 0.0}
\definecolor{amber}{rgb}{1.0, 0.75, 0.0}

\begin{document}

\title{Limitations of Stochastic Geometry Modelling for Estimating the Performance of CSMA Networks}

\author{Andra M. Voicu, Ljiljana Simi\'c, and Marina Petrova
\thanks{A. M. Voicu and L. Simi\'c are with the Institute for Networked Systems, RWTH Aachen University, 52072 Aachen, Germany (e-mail: \mbox{avo@inets.rwth-aachen.de}, \mbox{lsi@inets.rwth-aachen.de}).}
\thanks{M. Petrova is with the School of Electrical Engineering and Computer Science, KTH Royal Institute of Technology, 10044 Stockholm, Sweden (e-mail: \mbox{petrovam@kth.se})}
}

\markboth{ }%
{Voicu \MakeLowercase{\textit{et al.}}: Title}
%

\maketitle

\begin{abstract}

This letter considers stochastic geometry modelling (SGM) for estimating the signal-to-interference-and-noise ratio (SINR) and throughput of CSMA networks. We show that, despite its compact mathematical formulation, SGM has serious limitations in terms of both accuracy and computational efficiency. SGM often severely underestimates the SINR versus \mbox{ns-3} simulations, yet as it neglects the sensing overhead when mapping SINR to throughput, SGM usually overestimates the throughput substantially. We propose our \emph{hybrid} model for CSMA, which we argue is a superior modelling approach due to being significantly more accurate and at least one order of magnitude faster to compute than SGM.

\end{abstract}


\section{Introduction}

The increasing number of wireless devices and technologies operating in the unlicensed bands has lead to the wide adoption of CSMA at the MAC layer, especially since this is mandated as listen-before-talk (LBT) by some spectrum regulators, e.g. in the 5~GHz band in Europe.  
Representative wireless technologies that standardized variants of CSMA are the ubiquitous \mbox{Wi-Fi}, which implements CSMA/CA, but also newer technologies such as Licensed Assisted Access (LAA) and MulteFire, which extend LTE operation to the 5~GHz unlicensed band. 
With so many existing and emerging wireless deployments sharing the channel based on CSMA, accurately evaluating their performance and tuning their coexistence behaviour is of paramount importance. 

Two widely used performance metrics in this context are the signal-to-interference-and-noise ratio (SINR) and the throughput. 
Since it is often impractical to measure such metrics in real-world deployments or well-designed testbeds, various modelling approaches are used instead to estimate the CSMA network performance. 
Detailed discrete-event network simulators are employed to faithfully capture the performance of real deployments, e.g. \mbox{ns-3}, which has a full-stack protocol implementation and fine per-packet time granularity, but is computationally demanding for large deployments.  
By contrast, analytical models estimate the performance at a higher level of abstraction and are expected to be time-efficient to compute. 
Analytical models such as Bianchi's focus on the throughput per link without capturing interactions between nodes located outside the sensing range~\cite{Bianchi2000}, whereas stochastic geometry modelling (SGM) has emerged as a promising approach to estimate the SINR, throughput, or other interference and contention metrics for large-scale CSMA deployments~\cite{Nguyen2007, Li2016, Alfano2014, ElSawy2013, Haenggi2011}.
SGM assumes Poisson point process (PPP) node locations and estimates the average SINR and throughput distributions over the entire network, by means of compact mathematical formulations. 
Despite the popularity of SGM for CSMA, very little validation has been conducted since it was first proposed. 
The authors in e.g.~\cite{Li2016, Alfano2014, ElSawy2013} merely validate the correctness of their SGM formulations numerically, without any validation against more realistic CSMA models or simulations.
In~\cite{Nguyen2007} a partial validation of SGM against \mbox{ns-2} is presented, but only with respect to preliminary metrics, i.e. proportion of simultaneously transmitting access points (APs) and SINR coverage probability for a given AP-user distance. Consequently, it is still not clear how reliable the SINR and throughput estimates are, which are most relevant for estimating CSMA performance. 

In this letter we compare the SINR and throughput obtained by SGM~\cite{Li2016} against \mbox{Wi-Fi} simulations with \mbox{ns-3}.
We show that, despite its compact mathematical formulation, SGM has serious limitations in terms of accuracy and computational efficiency. 
We enhance SGM to incorporate the CSMA sensing overhead and thereby better estimate the throughput, but this cannot always compensate for the underlying inaccurate SINR estimation.  
Finally, we propose our so-called \emph{hybrid} model, which yields the time-average per-link SINR and throughput values over the CSMA network, by incorporating the specific path loss between each pair of nodes and Bianchi's approach for estimating the sensing overhead per node. We demonstrate that our hybrid model outperforms SGM in terms of accuracy and computational efficiency. We therefore argue it is superior to SGM for estimating the performance of CSMA networks. 
  

\section{Stochastic Geometry Model}

We consider SGM as initially proposed for CSMA~\cite{Nguyen2007} and subsequently extended in~\cite{Li2016}, also with explicit formulas for the SINR and throughput distributions. SGM in~\cite{Li2016} assumes two coexisting AP populations with different LBT (i.e. CSMA) and duty cycle MAC configurations. In this letter we show the underlying limitations of this model for the fundamental, single-population CSMA case.
We focus on the link SINR and throughput, for which SGM estimates distributions that characterize the overall network as an ensemble. The model assumes APs with a density of $\lambda_W$, distributed according to a PPP over an infinite area. The users are also PPP distributed and they associate with the closest AP, assuming a single user per AP. 
The performance metrics are mathematically derived by assuming a log-distance path loss model, Rayleigh fading, co-channel transmissions, and downlink saturated traffic.  
The model first defines the \emph{medium access probability of a typical AP}, $p^W_{0,MAP}(\lambda_W, 0)$, which is the probability that an AP is selected to transmit via the CSMA backoff procedure.
A related metric is the \emph{medium access probability of the tagged AP}, $\hat{p}^W_{0,MAP}(\lambda_W,0)$, i.e. the closest AP from the point of view of a user and thus the AP that the user associates with.
For the sake of brevity we omit the formal mathematical definitions, given in~\cite{Li2016}.
The SINR and the throughput metrics are further derived based on these two medium access probabilities. 

The \emph{SINR coverage probability} is the probability that the SINR of the user is strictly higher than a threshold $T$:  
\begin{equation}
\label{eq_sinrStoch_m}
\begin{split}
p^W_0&(T, \lambda_W,0) \approx \int\limits_{0}^{\infty} exp \Bigg(-\mu T l(r_0)\frac{N_0}{P_W}\Bigg) \\
                 & \times exp\Bigg(-\int\limits_{\mathbb{R}^2 \setminus B(0,r_0)} \frac{T l(r_0) \lambda_W h_1(r_0,x)}{l(||x||)+Tl(r_0)} dx\Bigg) f_W(r_0)dr_0,
\end{split}
\end{equation}
where $l(r_0)$ is the log-distance path loss, the Rayleigh fading parameter is $\mu$=1, $N_0$ is the noise power, $P_W$ is the transmit power level, $h_1(r_0,x)$ is the probability that other APs transmit given that the tagged AP transmits, ($r_0$, 0) are the polar coordinates of the tagged AP, $f_W(r_0)$ is the probability density function of the distance between a user and its tagged AP, and $B(0,r_0)$ is the open ball at the origin with radius $r_0$.

The \emph{rate coverage probability} with rate threshold $\rho$ is the probability for the tagged AP to support at least rate $\rho$: 
\begin{equation}
\label{eq_rateCov_m}
\begin{split}
P^W_{1,rate}(\lambda_W, 0, \rho, 1) = p^W_0(2^{\frac{\rho}{B\times \hat{p}^W_{0,MAP}(\lambda_W,0)}}-1, \lambda_W,0),
\end{split}
\end{equation}
where $B$ is the bandwidth.
The rate coverage probability is in fact the SINR coverage probability with threshold \mbox{$T=\left\{2^{\frac{\rho}{B\times \hat{p}^W_{0,MAP}(\lambda_W,0)}}-1\right\}$}, which depends on the Shannon capacity and the medium access probability $\hat{p}^W_{0,MAP}(\lambda_W,0)$, i.e. the fraction of transmit opportunities. 
Since the rate $\rho$ is first divided by $\hat{p}^W_{0,MAP}(\lambda_W,0)$ before being mapped to an SINR value, $\rho$ is an estimation of the throughput at the MAC layer, rather than the PHY rate, so we will refer to it as \emph{throughput}.
Importantly, in order to estimate the throughput coverage probability in (\ref{eq_rateCov_m}), we map the SINR to PHY rate with the \mbox{auto-rate} function of IEEE 802.11ac~\cite{IEEE2016} instead of the Shannon bound, for the sake of more realistic results and for consistency with the mapping in \mbox{ns-3}.
The SINR threshold in (\ref{eq_rateCov_m}) thus becomes $T=\rho^{-1}_x(\frac{\rho}{\hat{p}^W_{0,MAP}(\lambda_W,0)})$, where $\rho^{-1}_x(\cdot)$ is the inverse of the IEEE 802.11ac auto-rate function. 

By computing (\ref{eq_sinrStoch_m}) and (\ref{eq_rateCov_m}) for several values of $T$ and $\rho$, we obtain the complementary cumulative distribution functions (CCDFs) of the link SINR and throughput, respectively. We emphasize that these distributions characterize the links over the entire network and metrics specific to a given individual link cannot be obtained from SGM. 
We implement SGM computation in Matlab, where we use trapezoidal approximations of the function inside the integral in~(\ref{eq_sinrStoch_m}), due to singularities. 

\section{Proposed Models \& Enhancements}
\paragraph*{\textbf{Hybrid Model}}

We propose the \emph{hybrid} model for large-scale CSMA networks, so called since (i)~it takes into account the specific path losses between each pair of nodes, to estimate the time-average per-link SINR and throughput, and (ii)~incorporates Bianchi's model~\cite{Bianchi2000} to estimate the sensing overhead specific to each node. 
Our hybrid model thus captures diverse node location distributions and path loss models, has a finer time-average per-link granularity, and incorporates the inherent sensing overhead; by contrast, SGM captures only PPP node distributions and the log-distance path loss model, yields per-network metrics, and neglects the sensing overhead.  

The hybrid model is ideally suited for Monte Carlo simulations and we introduced a first, simpler version in~\cite{Voicu2016} for CSMA deployments coexisting with duty cycle deployments, for downlink saturated traffic, and one user per AP. 
In this letter we propose an enhanced hybrid model which now incorporates a major feature, namely the individual channel occupation time of each AP within the sensing range. This is important since it captures the different transmission durations of the APs and the interactions among these transmissions. Our hybrid model thus estimates with higher fidelity the AP throughput for adaptive modulation and coding (MCS) and different packet sizes, which lead to different transmission durations. 
We consider only the CSMA MAC and co-channel transmissions, consistent with SGM. 
The hybrid model assumes that APs within sensing range of each other share the channel in time via CSMA/CA, whereas APs outside the sensing range cause SINR-decreasing interference. APs are declared within sensing range if the received power from them is above the carrier sense threshold (CST). We present the hybrid model as follows.

The downlink time-average SINR is estimated for each individual user $u$ associated with an AP $x$ as
\begin{equation}\label{eq_29}
SINR_u=\frac{P_W/L_{u,x}}{ I_u + N_0 },
\end{equation}
where $L_{u,x}$ is the path loss between $x$ and $u$, and $I_u$ is the interference at $u$. 
For $I_u$ it is assumed that the interfering APs are located outside the sensing range and transmit for a fraction of time equal to the inverse of the number of APs within their respective sensing ranges:   
\begin{equation}
I_u = \displaystyle \sum_{z\in{\mathbf{A}\smallsetminus\mathbf{A}_x}} \frac{P_W/L_{u,z}}{1+|\mathbf{A}_z|}, 
\end{equation}
where $z$ is another AP, $\mathbf{A}$ is the set of all APs, $\mathbf{A}_x$ is the set of APs in $x$'s sensing range, $L_{u,z}$ is the path loss between $u$ and $z$, and $|\mathbf{A}_z|$ is the number of APs within $z$'s sensing range. 

We model the time-average per-link throughput of AP $x$ as
\begin{equation}\label{eq_28}
R_x=S_x \times AirTime_x \times \rho_x(SINR_u),
\end{equation}
\noindent where $S_x$ is the MAC efficiency of $x$ due to the sensing time based on Bianchi's model~\cite{Bianchi2000}, $AirTime_x$ is the average fraction of transmission time that $x$ occupies according to CSMA and the transmission durations in its sensing range, and $\rho_x(SINR_u)$ is the \mbox{auto-rate} function mapping the SINR of the associated user $u$ to the PHY rate of IEEE 802.11ac~\cite{IEEE2016}. 
The term \mbox{$AirTime_x=\frac{T_{f,x}\times  p_x}{T_{f,x} p_x + \displaystyle \sum_{z\in{\mathbf{A}_x}} T_{f,z}  p_z}$}, where $p_{x}= \frac{1}{1+|\mathbf{A}_{x}|}$ is the transmission probability of $x$, $T_{f,x}=PHY_{header}+\frac{MAC_{header}+MSDU}{\rho_x(SINR_u)}$ is the frame duration of $x$, and $p_z$ and $T_{f,z}$ are defined analogously. 
We emphasize that this is a major enhancement compared to our model in~\cite{Voicu2016}, which adopted the simpler approach, as also taken by SGM, of assuming equal transmission durations for all APs within sensing range, i.e. $AirTime_x= \frac{1}{1+|\mathbf{A}_x|}$.
Our hybrid model now estimates the AP throughput more realistically, since different transmission durations are typically observed in practical contemporary deployments, due to adaptive MCSs and different packet sizes.  

We define the MAC efficiency, $S_x$, as follows.
We model the MAC overhead due to sensing time based on the parameters of IEEE 802.11ac without RTS/CTS, by extending Bianchi's model~\cite{Bianchi2000} to incorporate different transmission durations of APs within sensing range~\cite{Voicu2016}. 
For each AP $x$ we estimate $S_x$:
\begin{equation}\label{eq_19}
S_x=\frac{\overline{T_{f,x}}}{\overline{T_{s,x}}-\overline{T_{c,x}}+\sigma\frac{\overline{T^*_{c,x}}-(1-\tau)^n(\overline{T^*_{c,x}}-1)}{n\tau(1-\tau)^{n-1}}},
\end{equation}
where $\overline{T_{f,x}}$ is the average duration of a frame in the sensing range of $x$, $\overline{T_{s,x}}$ is the average time the channel is occupied by a successful transmission in the sensing range of $x$, $\overline{T_{c,x}}$ is the average time the channel is occupied by a collision in the sensing range of $x$, $\sigma$=9~$\mu$s is the duration of an empty backoff time slot, $\overline{T^*_{c,x}}=\overline{T_{c,x}}/\sigma$, $n=1+|\mathbf{A}_x|$ is the total number of APs within the sensing range of AP $x$, and $\tau$ is the probability that a station transmits in a randomly chosen time slot. We calculate $\tau$ for binary exponential backoff with $CW_{min}$=15 and $CW_{max}$=1023~\cite{IEEE2016}.
The terms $\overline{T_{f,x}}$, $\overline{T_{s,x}}$, and $\overline{T_{c,x}}$ are defined in~\cite{Voicu2016} as functions of $\rho_x(SINR_u)$, $MSDU$=1500~bytes, $SIFS$=16~$\mu$s, $DIFS=SIFS+2\times\sigma$=34~$\mu$s, $ACK$=112~bits, $PHY_{header}$=40~$\mu$s, and $MAC_{header}$=320~bits (including FCS)~\cite{IEEE2016}.
We implement the hybrid model in Matlab.  

%

\paragraph*{\textbf{Enhanced SGM}}
Despite the sensing overhead being an intrinsic aspect of CSMA, SGM does not capture it. 
For a fair comparison with our hybrid model, we modify SGM to incorporate the sensing overhead similarly to the hybrid model. We introduce an average term $\overline{S_x}$ in (\ref{eq_rateCov_m}), based on Bianchi's model and similar to $S_x$ in (\ref{eq_19}). The SINR threshold in (\ref{eq_rateCov_m}) becomes $T=\left\{2^{\frac{\rho}{B\times \hat{p}^W_{0,MAP}(\lambda_W,0)  \times \overline{S_x} }}-1\right\}$ based on Shannon's capacity and $T=\rho^{-1}_x(\frac{\rho}{\hat{p}^W_{0,MAP}(\lambda_W,0)\times \overline{S_x}})$ with the IEEE 802.11ac auto-rate function. Importantly, for SGM the parameters specific to a given link cannot be calculated, so $\overline{S_x}$ is an average over all links in the network, where the durations $\overline{T_{f/s/c,x}}$ are the same for all APs and are estimated based on the PHY rate mapped to the median SINR over the entire network distribution. The average number of nodes within sensing range is $n=\frac{1}{p^W_{0,MAP}(\lambda_W, 0)}$.

\section{Evaluation Scenario \& Parameters}

We compare SGM, enhanced SGM, and the hybrid model against \mbox{ns-3}, in terms of SINR and throughput accuracy, and computational efficiency.
We select \mbox{ns-3} simulations as the ground-truth reference of realistic CSMA network performance, which we validated against \mbox{Wi-Fi} testbed measurements in~\cite{Voicu2017}. 
We assume co-channel transmissions over 20~MHz in the 5~GHz band, downlink saturated traffic, and one user per AP.
For all models, $P_W$=23~dBm, $NF$=15~dB, and $N_0$=\mbox{--174}~dBm/Hz.
 
For consistency with SGM, we consider a random PPP deployment scenario with a log-distance path loss with exponent $\alpha$=4~\cite{Li2016}. We assume realistic \mbox{Wi-Fi} network densities ranging from 500--10,000~APs/km\textsuperscript{2} and different CSTs between \mbox{--82}~dBm and \mbox{--62}~dBm, corresponding to values used in practice by \mbox{Wi-Fi} and LAA~\cite{Voicu2016}.
For \mbox{ns-3} and the hybrid model we define a total square area of 0.05~km\textsuperscript{2} over which APs and users are deployed and we analyse the results for only an inner central area of $\frac{1}{9}$ of the total area, to eliminate edge effects and thus model the equivalent of an infinite area as for SGM.  

For \mbox{ns-3} and the hybrid model we run Monte Carlo simulations with 50~realizations and we match the parameters as closely as possible. The node locations and the respective path losses between each pair of nodes are pre-computed and then fed to the simulations. 
Since we are interested in time-average SINR and throughput results, where SGM assumes Rayleigh fading with zero mean (i.e. $\mu$=1), we do not consider fast fading for \mbox{ns-3} and the hybrid model; fast fading would only increase the number of required Monte Carlo realizations to obtain equivalent average results to those from SGM.
We use the \mbox{Wi-Fi} module of \mbox{ns-3} and we average the SINR of every successfully received data packet for each link. The \mbox{ns-3} throughput is estimated based on the number of successfully received data packets that are forwarded by the MAC to the upper layer and traffic is simulated for 10~s.

\section{Results \& Discussion}

\begin{figure}[!t]
\centering
\includegraphics[width=0.95\columnwidth]{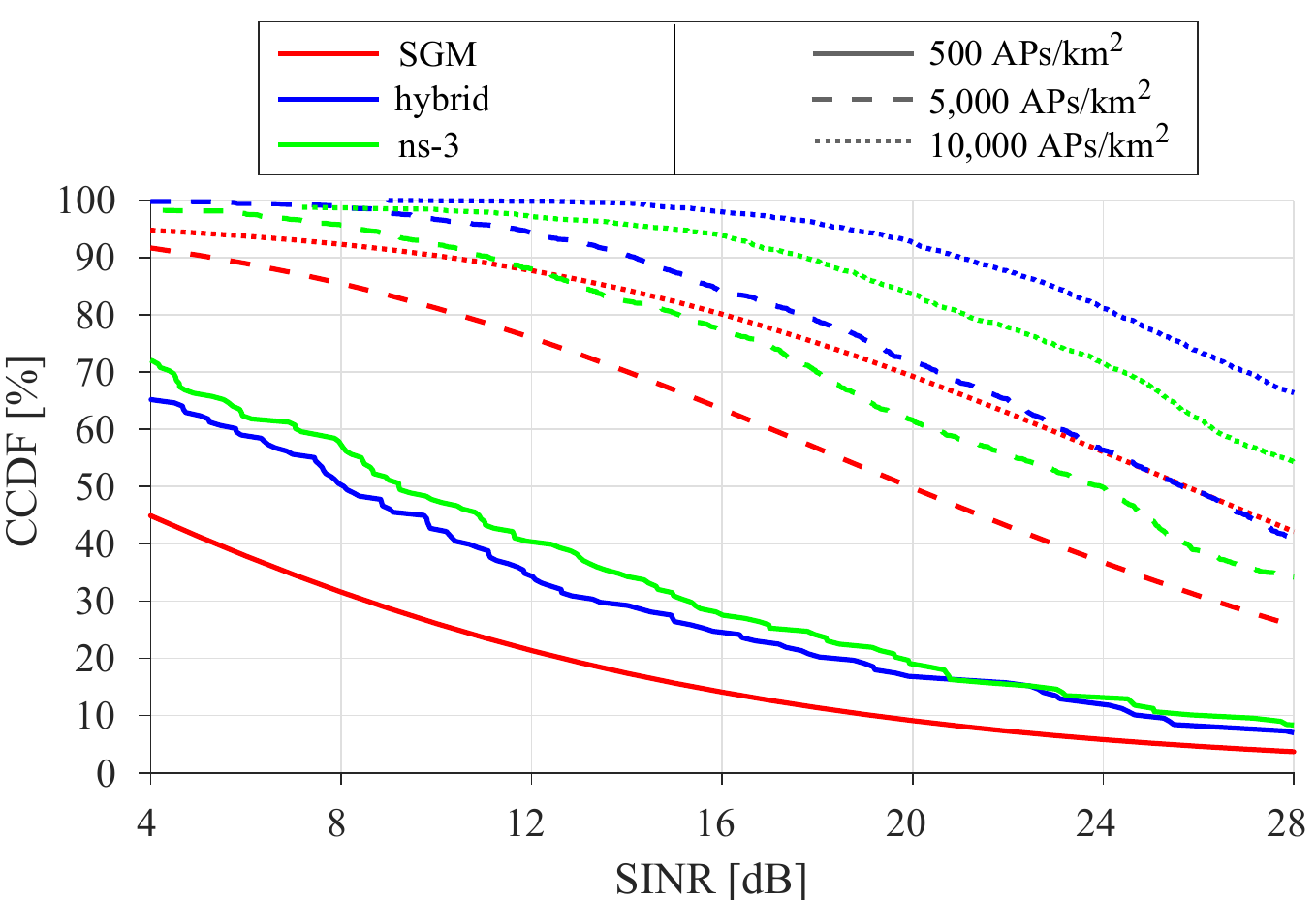} 
\caption{SINR for different AP densities and CST=--82~dBm.}
\label{fig_sinr_ppp_82}
\end{figure}

\paragraph*{\textbf{SINR Results}}
Fig.~\ref{fig_sinr_ppp_82} shows the SINR distribution for different network densities and CST=\mbox{--82}~dBm as for \mbox{Wi-Fi}. 
The relevant range of SINR is \mbox{\{4--27\}}~dB, where a minimum of 4~dB is required to establish a \mbox{Wi-Fi} link and the level of 27~dB corresponds to the highest IEEE 802.11ac PHY rate~\cite{IEEE2016}. 
The results confirm that SGM estimates a lower bound of the SINR from \mbox{ns-3} for all AP densities, as expected~\cite{Li2016}.
However, it is critical to also gauge by what margin the SINR is underestimated, in order to ensure that the results are still meaningful. For the density of 500~APs/km\textsuperscript{2}, 70\% of the APs establish a link in \mbox{ns-3}, whereas with SGM only 45\% of the APs have a link with their users. 
This difference of 25~percent points (pp) is significant and shows that SGM misleadingly suggests coverage problems in the network. 
For the higher SINR range or higher AP densities, the difference between SGM and \mbox{ns-3} remains non-negligible, but is reduced.
This occurs since, for low densities, the AP-user distance can be large, such that the impact of aggregate interference from APs outside the sensing range is greater than for high densities. This aggregate interference is overestimated by SGM due to approximating the interferer distribution and neglecting the sensing overhead of the interferers; in \mbox{ns-3}, as in practice, it is less likely that different interferers have random overlapping transmissions in time.    
In contrast to SGM, our hybrid model overestimates the SINR for moderate to large densities above 5,000~APs/km\textsuperscript{2} and underestimates the SINR for a low density of 500~APs/km\textsuperscript{2} with respect to \mbox{ns-3}. This results from the way that the hybrid model averages in time the interference from each node outside the sensing range, namely by dividing it by the number of nodes in the sensing range of the interferer. For low AP densities this overestimates the interference, since \mbox{ns-3} simulates also the sensing time of the interferers, so that fewer transmissions occur simultaneously, whereas for high AP densities, the more likely random overlapping transmissions become dominant in \mbox{ns-3} and exceed the average interference level estimated by the hybrid model.
Nonetheless, the hybrid model always estimates the SINR more accurately than SGM, within 10~pp of \mbox{ns-3}. 

\begin{figure}[!t]
\centering
\includegraphics[width=0.95\columnwidth]{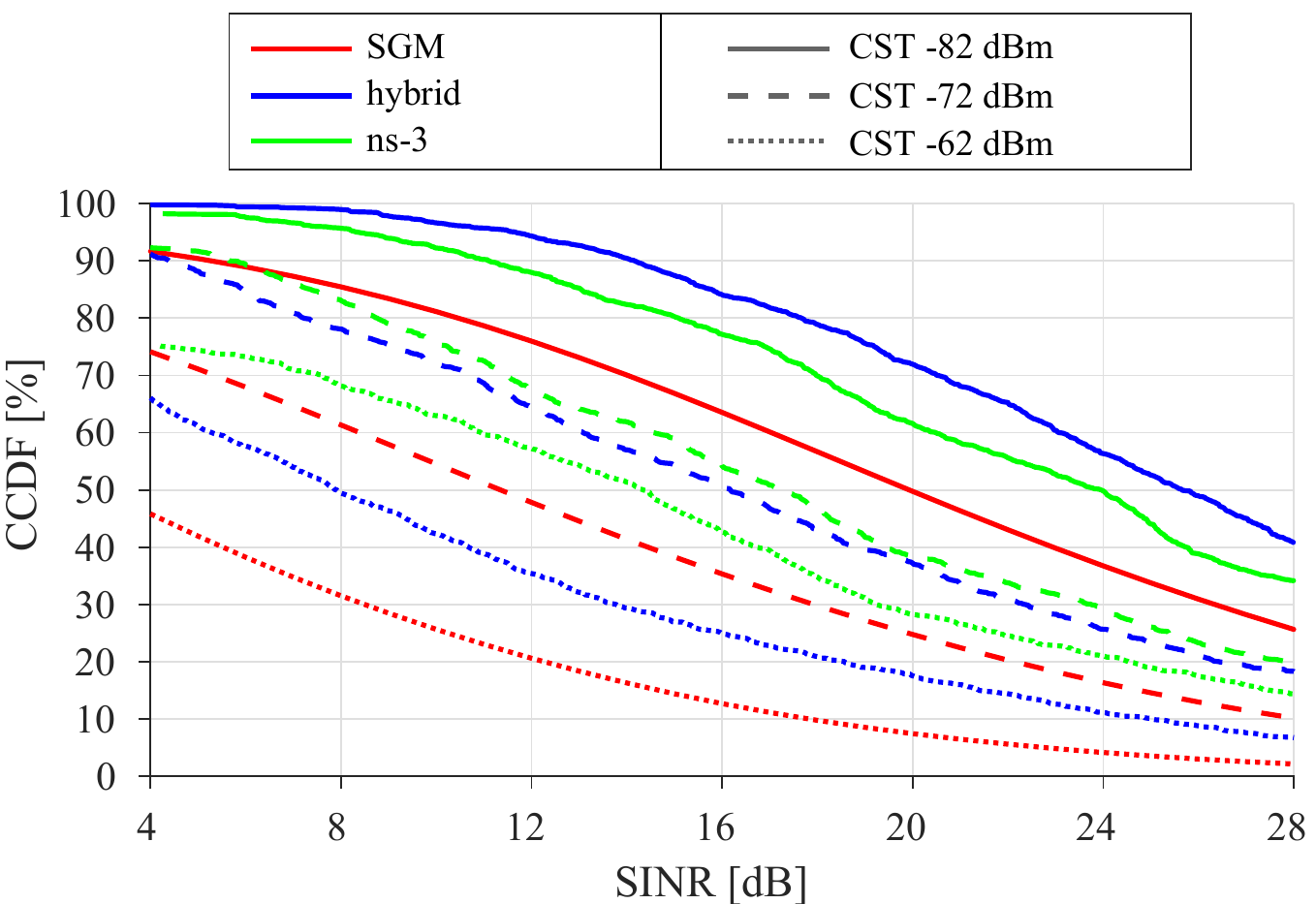}
\caption{SINR for different CSTs and 5,000 APs/km\textsuperscript{2}.}
\label{fig_sinr_ppp_diffCs}
\end{figure}

Fig.~\ref{fig_sinr_ppp_diffCs} shows the SINR distributions for 5,000~APs/km\textsuperscript{2} and different CSTs.
The results confirm that SGM always underestimates the SINR, also for CSTs other than \mbox{--82}~dBm.
The mismatch between SGM and \mbox{ns-3} increases up to 35~pp when the CST increases to \mbox{--62}~dBm, since for high CSTs there are more interferers outside the sensing range and thus they have a dominant impact over that of contending nodes within sensing range.  
Consistent with our discussion for Fig.~\ref{fig_sinr_ppp_82}, if the impact of aggregate interference is greater, the SINR mismatch between SGM and \mbox{ns-3} also increases.   
This shows that SGM estimates the SINR inaccurately for high CSTs currently applied in practice, e.g. for energy detection in \mbox{Wi-Fi}.
Our hybrid model overestimates the SINR compared to \mbox{ns-3} for a low CST and underestimates the SINR for high CSTs. Since the hybrid model averages the interference in time per interfering node, this results in lower aggregate interference than for \mbox{ns-3} if there are fewer and weaker interferers (low CST), but in higher aggregate interference if there are more and stronger interferers (high CST).   
Nonetheless, the hybrid model always yields a more accurate SINR than SGM, within 20~pp of \mbox{ns-3}.

\begin{figure}[!t]
\centering
\includegraphics[width=0.99\columnwidth]{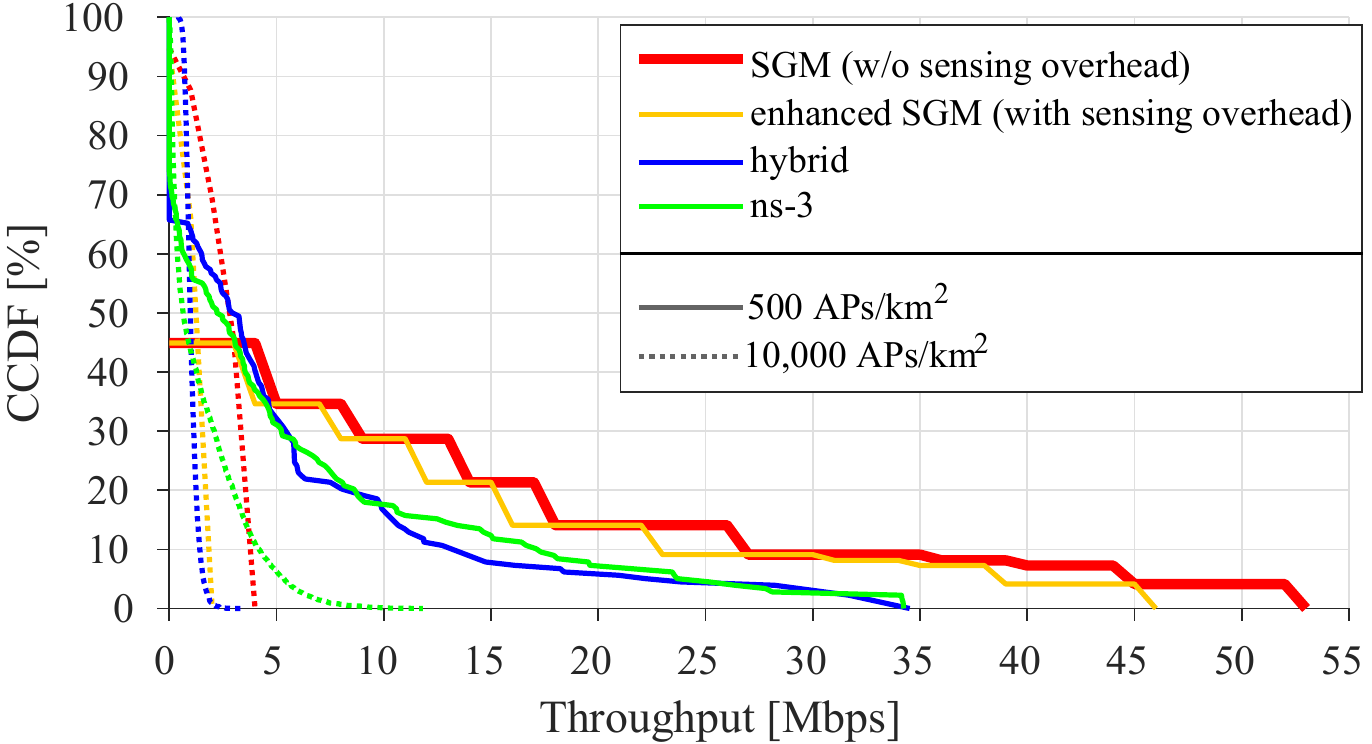}
\caption{Throughput for different AP densities and CST=\mbox{--82}~dBm.}
\label{fig_throughput}
\end{figure} 

\paragraph*{\textbf{Throughput Results}}

Fig.~\ref{fig_throughput} shows the link throughput distribution for different network densities and CST=\mbox{--82}~dBm, corresponding to the SINR in Fig.~\ref{fig_sinr_ppp_82}. 
Let us first consider SGM as originally proposed in~\cite{Li2016}, i.e. without MAC sensing overhead. 
We observe switching points between the throughput for SGM and \mbox{ns-3}, e.g. at 3~Mbps for 500~APs/km\textsuperscript{2}. Below this switching point SGM significantly underestimates the throughput, whereas above it, SGM significantly overestimates the throughput. The maximum throughput from SGM is 52~Mbps, i.e. 53\% higher than the maximum throughput from \mbox{ns-3}.  
Since SGM always underestimates the SINR (\emph{cf.} Fig.~\ref{fig_sinr_ppp_82}), it follows that: (i)~SGM underestimates the throughput, if the SINR is significantly underestimated; and (ii)~SGM overestimates the throughput, if the SINR is moderately underestimated and this is overcompensated by neglecting the sensing overhead.   
For the higher density of 10,000~APs/km\textsuperscript{2}, this switching point is shifted to a much lower throughput value, since SGM estimates the SINR more accurately at high densities. For this density we also observe a second switching point at 4~Mbps. This is the effect of applying a single average medium access probability and the auto-rate step function of IEEE~802.11ac to map the SINR to throughput for SGM. By contrast, \mbox{ns-3} counts the actual received packets for every user, so it has a finer time granularity and thus the throughput distribution forms a smoother curve.  
 
Let us now consider SGM with our enhancement for sensing overhead. 
Fig.~\ref{fig_throughput} shows that, the enhanced SGM estimates a more accurate throughput than SGM without the sensing overhead for 10,000~APs/km\textsuperscript{2}. However, for 500~APs/km\textsuperscript{2}, the sensing overhead cannot sufficiently compensate for the SINR underestimation. Specifically, SGM incorrectly estimates a low median SINR, which is then mapped to a low PHY rate, thus a long transmission duration and in turn a small sensing overhead. As such, the sensing overhead decreases the original throughput only marginally.
This shows overall that enhancing SGM to incorporate the sensing overhead according to Bianchi's model yields a more accurate throughput only if the SINR is not severely underestimated; large inaccuracies in SINR cannot be compensated for, so the corresponding throughput estimate remains inaccurate.    

Finally, Fig.~\ref{fig_throughput} shows that the throughput estimated by the hybrid model is accurate with respect to \mbox{ns-3} for 500~APs/km\textsuperscript{2}. This is owing to both estimating well the SINR in Fig.~\ref{fig_sinr_ppp_82} and incorporating the sensing overhead through $S_x$ in (\ref{eq_19}). For the higher network density of 10,000~APs/km\textsuperscript{2}, the hybrid model becomes more sensitive to averaging the interference, air time, and sensing overhead per link and thus estimates a less accurate throughput. 
In fact, for this density, the hybrid model yields similar results as enhanced SGM. Nonetheless, the hybrid model yields overall a closer throughput estimate to \mbox{ns-3} than SGM.            

\paragraph*{\textbf{Computation Time}}

\begin{figure}[!t]
\centering
\subfloat[stochastic geometry]{\includegraphics[width=0.95\columnwidth]{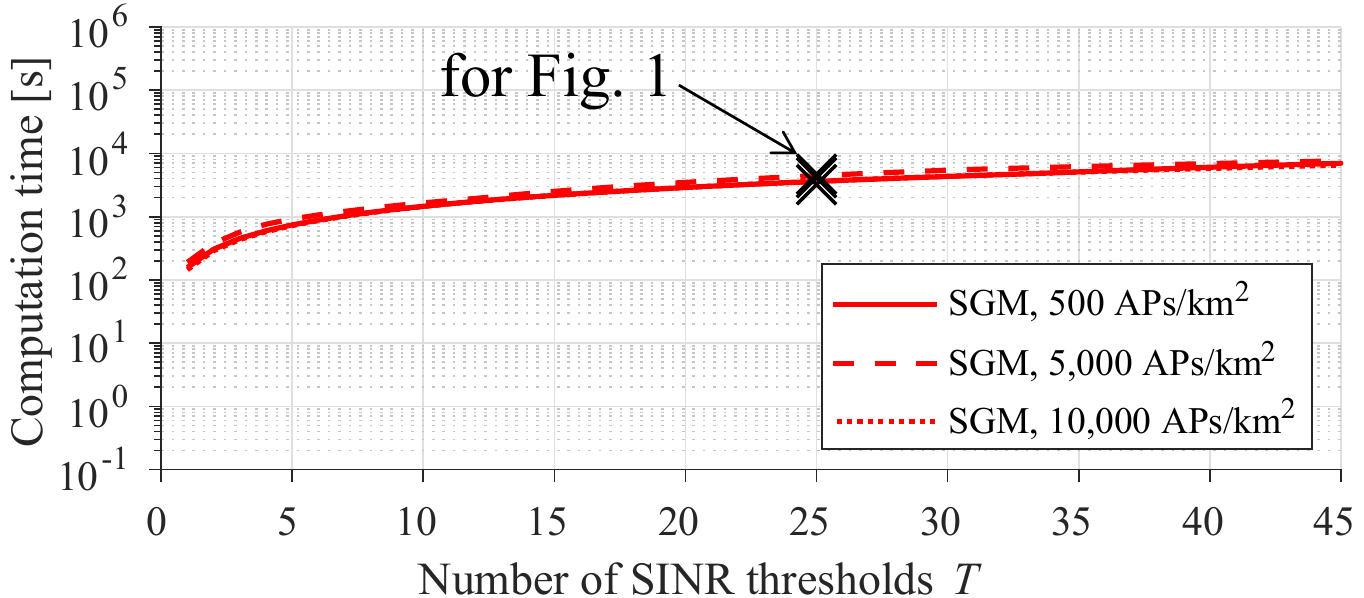} \label{fig_computTime_1}}
\\
\subfloat[ns-3 and hybrid]{\includegraphics[width=0.95\columnwidth]{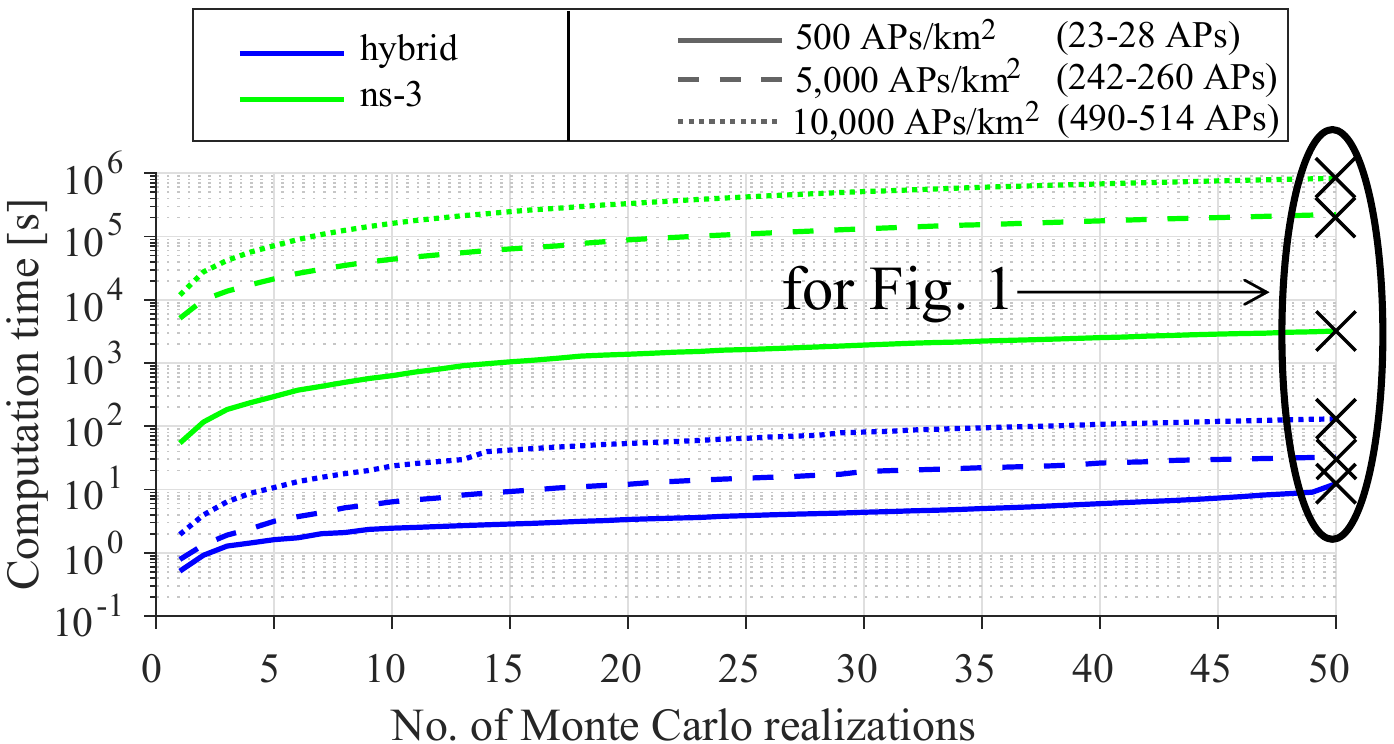} \label{fig_computTime_2}}
\caption{Computation time for different network densities and CST=\mbox{--82}~dBm.}
\label{fig_computTime}
\end{figure}

Fig.~\ref{fig_computTime} shows the computation time per core required to obtain the SINR and throughput distributions. We note that all computations were run on high-performance simulation servers with Intel Skylake Platinum 8160 processors.
The computation time for SGM in Fig.~\ref{fig_computTime_1} increases linearly with the number of SINR thresholds $T$ for which (\ref{eq_sinrStoch_m}) is applied. The time for a single $T$ was at least 142~s and for obtaining the results in Fig.~\ref{fig_sinr_ppp_82}, the total computation time ranged between 60 and 73~minutes for each AP density. 
By contrast, the computation time of \mbox{ns-3} and the hybrid model in Fig.~\ref{fig_computTime_2} depends on the number of Monte Carlo network realizations and the number of deployed nodes.
For \mbox{ns-3}, the time for a single realization increased from 38~s to 5~hours as the number of deployed APs increased (i.e. higher AP densities). All realizations simulated to plot Fig.~\ref{fig_sinr_ppp_82} lasted 53~minutes to 232~hours for each AP density. The computation time of \mbox{ns-3} thus varies over a much wider range than for SGM, and for large numbers of nodes it is several orders of magnitude higher.    
Finally, we emphasize that the computation time for the hybrid model is significantly lower than both SGM and \mbox{ns-3}. The time for a single realization ranged between 71~ms and 10~s as the number of deployed APs increased with different AP densities.
For all realizations generated to plot Fig.~\ref{fig_sinr_ppp_82}, the hybrid model took between 13 and 131~s for each AP density. Importantly, for the hybrid model the computation time is at least one order of magnitude lower than for SGM, for obtaining corresponding SINR and throughput distributions.

\paragraph*{\textbf{Summary \& Findings}}
We showed that SGM estimates poorly the SINR of CSMA deployments with low AP densities or high CSTs. 
Furthermore, SGM either significantly underestimates, or significantly overestimates the throughput, due to poor estimation of the SINR and omission of the sensing overhead. Although for high AP densities we successfully incorporated the sensing overhead, the impact of SGM severely underestimating the SINR for low AP densities was not overcome.  
By contrast, our proposed hybrid model always estimates the SINR and the throughput more accurately than SGM, yields results with finer granularity as a time-average for each individual link, captures diverse node deployments and path loss models, while requiring a significantly lower computation time than SGM.

\section*{Acknowledgments}

Simulations were performed with computing resources granted by RWTH Aachen University under project rwth0352.

\bibliographystyle{IEEEtran}
\bibliography{IEEEabrv, bibliography}


\end{document}